\documentclass[sigconf,twocolumn]{acmart}
\AtBeginDocument{%
  \providecommand\BibTeX{{%
    \normalfont B\kern-0.5em{\scshape i\kern-0.25em b}\kern-0.8em\TeX}}}

\copyrightyear{2021}
\acmYear{2021}
\setcopyright{acmcopyright}\acmConference[VaMoS'21]{15th International Working Conference on Variability Modelling of Software-Intensive Systems}{February 9--11, 2021}{Krems, Austria}
\acmBooktitle{15th International Working Conference on Variability Modelling of Software-Intensive Systems (VaMoS'21), February 9--11, 2021, Krems, Austria}
\acmPrice{15.00}
\acmDOI{10.1145/3442391.3442408}
\acmISBN{978-1-4503-8824-5/21/02}

\begin{document}

\title{An Overview of Recommender Systems and Machine Learning
in Feature Modeling and Configuration}


\author{Alexander Felfernig}
\affiliation{
  \institution{Institute of Software Technology, Graz University of Technology}
  \city{Graz}
  \country{Austria}}
\email{alexander.felfernig@ist.tugraz.at}

\author{Viet-Man Le}
\affiliation{
  \institution{Institute of Software Technology, Graz University of Technology}
  \city{Graz}
  \country{Austria}}
  \email{vietman.le@ist.tugraz.at}

\author{Andrei Popescu}
\affiliation{
  \institution{Institute of Software Technology, Graz University of Technology}
  \city{Graz}
  \country{Austria}}
  \email{andrei.popescu@ist.tugraz.at}

\author{Mathias Uta}
\affiliation{
  \institution{Siemens  Gas  \&  Power}
  \city{Erlangen}
  \country{Germany}}
  \email{mathias.uta@siemens.com}

\author{Thi Ngoc Trang Tran}
\affiliation{
  \institution{Institute of Software Technology, Graz University of Technology}
  \city{Graz}
  \country{Austria}}
  \email{ttrang@ist.tugraz.at}

\author{Müslüm Atas}
\affiliation{
  \institution{Institute of Software Technology, Graz University of Technology}
  \city{Graz}
  \country{Austria}}
  \email{muatas@ist.tugraz.at}



\begin{abstract}
Recommender systems support decisions in various domains ranging from simple items such as books and movies to more complex items such as financial services, telecommunication equipment, and software systems. In this context, recommendations are determined, for example, on the basis of analyzing the preferences of similar users. In contrast to simple items which can be enumerated in an item catalog, complex items have to be represented on the basis of variability models (e.g., feature models) since a complete enumeration of all possible configurations is infeasible and would trigger significant performance issues. In this paper, we give an overview of a potential new line of research which is related to the application of recommender systems and machine learning techniques in feature modeling and configuration. In this context, we give  examples  of the application of recommender systems and machine learning and discuss  future research issues.
\end{abstract}

\maketitle

\section{Introduction}\label{introduction}

Feature models can be regarded as a central element of feature-oriented software development (FOSD) processes \cite{Apel2009}. Feature models can be used to represent variability and commonality properties of software artifacts and various other types of products and services  \cite{Acher2018,BeSeRu2010,Felfernigetal2014,Kang1990}. Applications thereof support users in deciding about which features should be included in a specific configuration. Feature models and variability models in general can become quite complex which makes it challenging to develop these models as well as to interact with the corresponding decision support systems \cite{Falkner2011,Pereira2018,RodasSilva2019}. In this paper, we give an overview of a potentially new research line which is related to the application of recommender systems and machine learning techniques in feature modeling and configuration scenarios.

Recommender systems can be defined \emph{as any system that guides a user in a personalized way to interesting or useful objects in a large space of possible options or that produces such objects as output} \cite{FelfernigBurke2008}. These systems use basic machine learning techniques (classification as well as prediction techniques) for being able to identify items of relevance for a user. Typical applications of recommender systems rely on a dataset that serves as an input for learning algorithms. These algorithms infer models that predict item preferences of users. Recommender system applications are manifold and range from simple items such as news  \cite{Konstan1997} to more complex items such as financial services \cite{Felfernig2006} and software systems \cite{RodasSilva2019}. 

In general, recommender systems help to infer user interests on the basis of preference definition histories, i.e., which items were preferred by users in the past. \emph{Collaborative filtering} recommender systems \cite{Konstan1997} are based on the idea of word-of-mouth promotion where items regarded as relevant by users with similar preferences (so-called \emph{nearest neighbors}) are recommended to the current user. A model-based variant thereof is \emph{matrix factorization} \cite{Koren2009} which describes the relationship between users and items on the basis of a set of hidden aspects.\footnote{In matrix factorization \cite{Koren2009}, these aspects are also denoted as \emph{features}.} \emph{Content-based filtering} \cite{Pazzani1997} is based on the idea of recommending items to the current user which are similar to those the user has liked in the past. \emph{Knowledge-based recommender systems} \cite{Burke2000} are based on explicit recommendation knowledge in terms of attributes and constraints or similarity metrics which describe the relationship between a set of customer requirements and corresponding items. Finally, \emph{group recommender systems} \cite{Felfernigetal2018} focus on the recommendation of items to groups of users instead of single users.

The major goal of this paper is to analyze recommendation scenarios in the context of feature model development and configuration. Feature models describe variability and commonality properties of items. In many cases, feature models are the basis of configurator applications associated with a potentially large user base. In application scenarios where user communities are interacting with configurators (derived from feature models), data can be collected from user interactions and exploited  to predict  user-individual preferences.

Summarizing, the contributions of this paper are the following:
\begin{itemize}
\item{We provide an overview of  recommendation and machine learning approaches in feature modeling and configuration. In this context, we focus on the basic scenarios of \emph{interactive configuration}, \emph{reconfiguration}, and \emph{feature modeling processes}.}
\item{On the basis of examples, we sketch application scenarios of recommendation technologies that open a new line of research in feature modeling and configuration.} 
\item{Finally, we discuss open research issues to be solved to further advance the related state of the art.}
\end{itemize}

The remainder of this paper is organized as follows. In Section \ref{featuremodelscsp}, we introduce an example feature model from the domain of \emph{survey software services}. In Section \ref{configurationranking}, we introduce a CSP-based representation of the example feature model. In Section \ref{configuration}, we discuss scenarios in which recommendation and related machine learning techniques can be applied to support \emph{interactive configuration}. Section \ref{reconfiguration} focuses on recommender systems and machine learning in the context of \emph{reconfiguration}. Finally, Section \ref{modeling} provides an overview of recommendation concepts that can be applied to support feature model \emph{knowledge acquisition}. Sections \ref{configuration}--\ref{modeling} also include a topic-specific discussion of open research issues. In Section \ref{conclusions}, the paper is concluded with a discussion of further research topics.

\section{An Example Feature Model}\label{featuremodelscsp}

Features can be organized in a hierarchical fashion \cite{BeSeRu2010} using relationships such as \emph{mandatory} (if a parent feature is included in a configuration, the child feature must be included as well, and vice versa), \emph{optional} (given the inclusion of a parent feature, the inclusion of the corresponding child feature is optional), \emph{alternative} (if the parent feature is included, exactly one of the child features has to be included), and \emph{or} (if the parent feature is included, at least one of the child features has to be included). Furthermore, cross-tree constraints can be used to define relationships between features that do not follow the hierarchical structure of the feature model. First, \emph{excludes(a,b)} constraints prohibit the inclusion of both features ($a$ and $b$) in the same configuration. Second, \emph{requires(a,b)} constraints necessitate that if feature $a$ is included in a configuration, feature $b$ must be included as well.

\begin{figure}[ht] 
			\centering
			\fbox{
				\includegraphics[width=0.45 \textwidth]{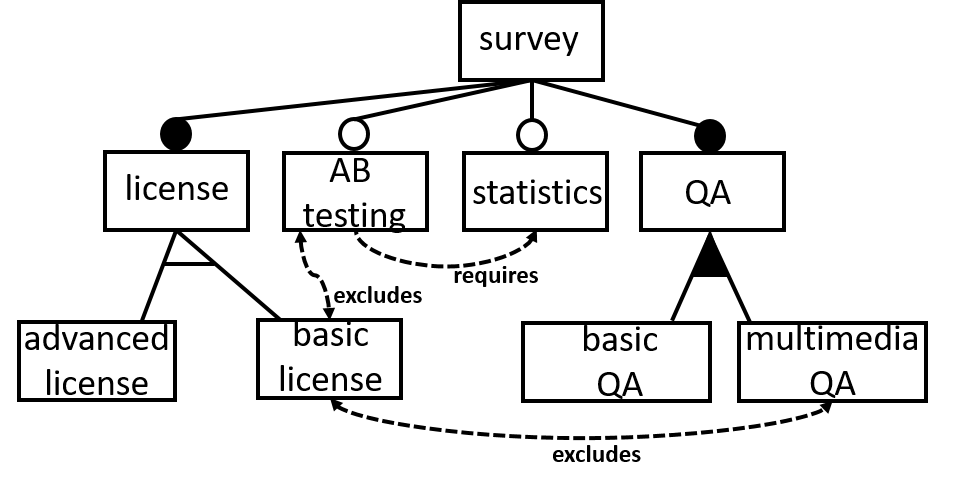}
			}
			\caption{An example feature model (\emph{survey software}).} \label{Figure1}
		\end{figure}

An example feature model is depicted in Figure \ref{Figure1}. In this  model, \emph{license} is used to describe the selected license model where two different models are available: an \emph{advanced license} allows to include all features provided in the feature model whereas a \emph{basic license} restricts the set of selectable features. In Figure \ref{Figure1}, \emph{license} and \emph{QA} are designed as mandatory features, i.e., must be included in every survey software configuration. If a user selects \emph{ABtesting} to be included in the configuration, this also requires the inclusion of the \emph{statistics} feature. The \emph{QA} feature supports both, \emph{basic QA} and \emph{multimedia QA} questions -- at least one of these has to be included in each configuration.

\section{Configuration Task and Solution Ranking}\label{configurationranking}	

To enable reasoning about potential solutions (configurations), a feature model has to be translated into a formal representation. One option for a formal representation of feature models are constraint satisfaction problems (CSPs) \cite{Tsang1993}. On the level of a CSP, a feature model can be defined as a \emph{configuration task} (see Definition \ref{def1}).

\definition{Configuration Task\label{def1}}. A configuration task derived from a feature model can be defined as a constraint satisfaction problem (CSP) $(V,D,C)$ where $V=\{v_1,v_2,..,v_n\}$ is a set of Boolean variables, $D=\{dom(v_1),$ $dom(v_2),..,dom(v_n)\}$ is the  set of variable domains, and $C=C_R \cup C_F$ where $C_R=\{c_1,c_2,..,c_k\}$ is a set of customer requirements (i.e., preferred inclusions and exclusions of features), and $C_F=\{c_{k+1},c_{k+2},..,c_q\}$ is a corresponding set of constraints derived from the feature model. 

In the case of basic feature models (without further attributes), $dom(v_i) = \{0,1\}$. Constraints $c_j$ can be directly derived from a feature model and represent (1) \emph{structural relationships} (e.g., a mandatory relationship) and (2) \emph {cross-tree relationships} (e.g., a \emph{requires} relationship). A set of rules of how to formalize the relationships of a feature model in terms of a set of corresponding constraints is discussed, for example, in Benavides et al. \cite{BeSeRu2010}.

On the basis of the definition of a \emph{configuration task}, we now introduce the definition of a \emph{configuration} (Definition \ref{def2}).

\definition{Configuration\label{def2}}. A configuration (solution) for a configuration task $(V,D,C)$ is an assignment $A$ of the variables in $V$ which fulfils the criteria that all constraints in $C$ are consistent with the variable assignments in $A$.

Following Definition \ref{def1}, Table \ref{tab:variablesanddomains} represents the set of CSP variables (i.e., features represented by variables in $V$) and corresponding Boolean domains ($D$) derived from the feature model in Figure \ref{Figure1}. 
	
\begin{table}[ht]
\centering 
\caption{Features (including abbreviations of feature names) and corresponding domain definitions ($1=true$, $0=false$).}
\begin{tabular}{|c|c|c|c|c|c|c|c|c|}
\hline
  featurename    & abbreviation                &  domain      
\tabularnewline
\hline \hline
$survey$ &$sur$                      &   $\{0,1\}$             
\tabularnewline
\hline
$license$ & $lic$                      &   $\{0,1\}$             
\tabularnewline
\hline
$advancedlicense$ & $adlic$               &   $\{0,1\}$                 
\tabularnewline
\hline
$basiclicense$ & $baslic$           &   $\{0,1\}$                
\tabularnewline
\hline
$ABtesting$ & $AB$                &   $\{0,1\}$                
\tabularnewline
\hline
$statistics$ & $stat$                &   $\{0,1\}$                
\tabularnewline
\hline
$QA$ & $QA$                &   $\{0,1\}$                
\tabularnewline
\hline
$basicQA$ & $basQA$       &   $\{0,1\}$              
\tabularnewline
\hline
$multimediaQA$ & $mmQA$            &   $\{0,1\}$               
\tabularnewline
\hline
\end{tabular} 
\label{tab:variablesanddomains}
\end{table}

In our example, Table \ref{tab:constraintscsp} shows the constraints $C_F$ derived from the feature model depicted in Figure \ref{Figure1}. 

\begin{table}[ht]
\centering 
\caption{Constraints $C_F=\{c_1..c_9\}$ derived from Figure \ref{Figure1}.}
\begin{tabular}{|c|c|c|c|c|c|c|c|c|}
\hline
  constraint                 &  CSP representation       
\tabularnewline
\hline \hline
$c_0$                      &   $survey=1$             
\tabularnewline
\hline
$c_1$                      &   $survey$ $\leftrightarrow license$             
\tabularnewline
\hline
$c_2$                    &   $ABtesting \rightarrow survey$             
\tabularnewline
\hline
$c_3$                  &   $statistics \rightarrow survey$            
\tabularnewline
\hline
$c_4$                  &   $survey \leftrightarrow QA$            
\tabularnewline
\hline
$c_5$                  &   $QA \rightarrow basicQA \lor multimediaQA$            
\tabularnewline
\hline
$c_6$           &   $(advancedlicense \leftrightarrow \neg basiclicense \land license) \land $ \tabularnewline  
                    &   $(basiclicense \leftrightarrow \neg advancedlicense \land license)$  
\tabularnewline
\hline
$c_7$            &   $\neg(ABtesting \land basiclicense)$            
\tabularnewline
\hline
$c_8$            &   $ABtesting \rightarrow statistics$            
\tabularnewline
\hline
$c_9$            &   $\neg(basiclicense \land multimediaQA)$            
\tabularnewline
\hline
\end{tabular} 
\label{tab:constraintscsp} 
\end{table}

In Table \ref{tab:constraintscsp}, $c_0$ represents a so-called \emph{root constraint} which is used to assure that (irrelevant) empty configurations are avoided. Furthermore, $c_1$ and $c_4$ represent mandatory relationships: both \emph{payment} and \emph{QA} are mandatory, i.e., every configuration has to include these features. Both of the features \emph{ABtesting} and \emph{statistics} are regarded as optional (constraints $c_2$ and $c_3$), i.e., they do not have to be part of every configuration. The features \emph{basiclicense} and \emph{advancedlicense} are regarded as part of an alternative relationship ($c_6$). Furthermore, \emph{basicQA} and \emph{multimediaQA} are part of an optional relationship ($c_5$). Finally, the features \emph{ABtesting} and \emph{basiclicense} are regarded as incompatible ($c_7$), the inclusion of feature \emph{ABtesting} requires the inclusion of feature \emph{statistics} ($c_8$), and the features \emph{basiclicense} and \emph{$multimediaQA$} are incompatible.

 Assuming the existence of the customer (user) requirement $C_R=\{c_{10}: ABtesting=1\}$, we are able to derive the configurations $A_i$ as depicted in Table \ref{tab:configurationexample}. All three configurations could be regarded as \emph{recommendation candidates}, however, we are primarily interested in solutions which are the most relevant ones for a user. In the following, we assume the existence of \emph{two example users}, namely $u_a$ and $u_b$ (both would like to have included the feature $ABtesting$).  Following the concepts of utility-based ranking which is a major element of knowledge-based recommendation \cite{Felfernig2006}, \emph{interest dimensions} can be regarded as explicit global solution properties. We denote a specific interest dimension-related \emph{user preference} of user $u_i$ with regard to dimension $d_j$ as $up(u_i,d_j)$, for example, $up(u_a,simplicity)=0.8$, $up(u_a,productivity)=0.2)$, $up(u_b,simplicity)=0.2)$, and $up(u_b,productivity)=0.8)$.
 
  \begin{table}[ht]
\centering 
\caption{Example configurations $A_1..A_3$ consistent with $C = C_F \cup C_R$ ($C_R=\{c_{10}: ABtesting=1\}$).}
\begin{tabular}{|c|c|c|c|c|c|c|c|c|}
\hline
  feature                 &  $A_1$            & $A_2$         & $A_3$ 
\tabularnewline
\hline \hline
$survey$                      &   1            &  1         &  1
\tabularnewline
\hline
$advancedlicense$               &   1            &  1         &  1
\tabularnewline
\hline
$basicicense$           &   0           &  0        &  0
\tabularnewline
\hline
$ABtesting$              &   1            &  1         &  1
\tabularnewline
\hline
$statistics$          &   1            &  1         &  1
\tabularnewline
\hline
$basicQA$         &   1            &  0        &  1
\tabularnewline
\hline
$multimediaQA$            &   0           &  1         &  1
\tabularnewline
\hline
\end{tabular} 
\label{tab:configurationexample} 
\end{table}

The higher the value of an interest dimension (between $0..1$), the higher the corresponding interest of the user. For example, if a user is interested in \emph{simplicity}, he or she will prefer configurations with a lower number of features (lower overhead in understanding the provided software). In order to be able to rank configurations, we also need to evaluate the solution (configuration) attributes with regard to the interest dimensions \emph{simplicity} and \emph{productivity}. In this context, we omit the features \emph{survey}, \emph{license}, and \emph{QA} which are included in every configuration.

\begin{table}[ht]
\centering 
\caption{Utility evaluation $u$ of features $f_i$  with regard to interest dimensions $d_j$ ($u(f_i,d_j)$), for example, $u(advancedlicense,simplicity)=0.1$.}
\begin{tabular}{|c|c|c|c|c|c|c|c|c|}
\hline
  features $f_i$           &  simplicity      & productivity
\tabularnewline
\hline \hline
$advancedlicense=1$                   &   0.1           & 1.0
\tabularnewline
\hline
$basiclicense=1$                 &   1.0          &  0.1   
\tabularnewline
\hline
$ABtesting=1$                 &   0.3           &  1.0
\tabularnewline
\hline
$statistics=1$                  &   0.5          &  1.0
\tabularnewline
\hline
$multimediaQA=1$         &   0.3          &  1.0
\tabularnewline
\hline
$basicQA=1$           &   1.0          &  1.0
\tabularnewline
\hline
\end{tabular} 
\label{tab:interestdimensionsattributes}
\end{table}

On the basis of the preferences of $u_a$ and $u_b$, and the information provided in Table \ref{tab:interestdimensionsattributes}, we are able to calculate the overall utility of the individual solutions $A_1..A_3$ using Formula \ref{eq:maututility}. The user-individual utilities are shown in Table \ref{tab:utilitycalculations}. In our simplified example, alternative $A_3$ is assumed to be the preferred configuration of both users. Different users could have different preferences and could also receive different recommendations that depend on their personal preferences regarding a set of interest dimensions.

\begin{equation}\label{eq:maututility}
utility(A,u_j)=\Sigma_{f=true \in A} \Sigma_{d \in Dims} u(f,d)\times up(u_j,d)
\end{equation}

\begin{table}[ht]
\centering 
\caption{Utilities of example configurations $A_1..A_3$ for users $u_a$ and $u_b$.}
\begin{tabular}{|c|c|c|c|c|c|c|c|c|}
\hline
configuration                 &  $A_1$            & $A_2$         & $A_3$ 
\tabularnewline
\hline \hline
$utility(A_i,u_a)$            &   1.76       &       2.32       &     \bf2.76        
\tabularnewline
\hline
$utility(A_i,u_b)$           &    3.44     &      3.58       &     \bf4.44     
\tabularnewline
\hline
\end{tabular} 
\label{tab:utilitycalculations}
\end{table}

Utility-based ranking (recommendation) can be used if \emph{alternative configurations have already been determined} \cite{Felfernig2006}. Note that we focused on a scenario where utility-based recommendation is used to identify a recommendation of relevance for a single user. However, there are also scenarios where recommendations have to be determined for \emph{groups of users} \cite{Felfernigetal2018}. In such a context, the preferences of individual users (group members) are aggregated, for example, by interpreting a group rating as the average value of the user-individual item evaluations. For details regarding the application of group recommender systems in configuration contexts we refer to \cite{Felfernigetal2016,Felfernigetal2018}.

Utility-based ranking has the disadvantage of \emph{knowledge acquisition efforts} that are needed to specify the contributions of user selections to individual interest dimensions. In the following section, we will discuss  scenarios in which recommender systems can be applied to support users in the selection of individual features, i.e., the configuration process is still ongoing and users need support in selecting and deselecting individual features.

\section{Supporting Interactive Configuration}\label{configuration}

Assuming the availability of user interaction data from previous configuration sessions, we are able to proactively support the \emph{current user} interacting with a configurator \cite{Falkner2011,Pereira2018,RodasSilva2019}. A need for such a functionality is given if a user has \emph{limited domain knowledge} and is unsure about the inclusion or exclusion of a specific feature or a user simply \emph{does not have the time to specify every feature} (which is often the case with large and complex feature models). Table \ref{tab:usersessionsselectedfeatures} depicts a simplified example of such a scenario where users ($u_1..u_3$) have already completed their configuration sessions. The $current$ user just started his/her session and has specified his/her preferences regarding the features \emph{lic} (\emph{license}), \emph{adlic} (\emph{advanced license}), and \emph{baslic} (\emph{basic license}). 

\begin{table}[ht]
\centering 
\caption{Recommendation of a feature value ($0$) for feature $ABtesting$ to the current user.}
\begin{tabular}{|c|c|c|c|c|c|c|c|c|}
\hline
  user/   &  lic  & adlic   & baslic     & AB    & stat          & QA      &bas-     &   mm-
\tabularnewline
 session   &    &    &      &     &           &       &  QA   &  QA 
\tabularnewline
\hline \hline
$u_1$           &   1   &  0    &  1        &  0    &     1         &  1        &   1   &    0
\tabularnewline
\hline
$u_2$           &   1   &  1    &  0        &  1    &     1         &  1        &   1   &    1
\tabularnewline
\hline
$u_3$           &   1   &  0    &  1        &  0    &     1         &  1        &   1   &    0
\tabularnewline
\hline
current     &   1   &  0    &  1        &  ?$\rightarrow$0    &     -         &  -        &   -   &    -
\tabularnewline
\hline
\end{tabular}
\label{tab:usersessionsselectedfeatures} 
\end{table}

The idea of \emph{collaborative filtering} based recommendation \cite{Konstan1997} is to analyze  users with similar preferences compared to the current user (\emph{nearest neighbors}) and exploit this knowledge for determining recommendations for the \emph{current user}.\footnote{For a detailed analysis of different collaborative filtering approaches, we refer to Ekstrand et al. \cite{Ekstrand2011}.} For example, the users with the most similar preferences compared to the current user are $u_1$ and $u_3$. If we are interested in recommending feature inclusion or exclusion for the feature $ABtesting$ ($AB$), we could recommend to the current user to follow his/her nearest neighbors $u_1$ and $u_3$, i.e., not to include this feature. In such scenarios, the number of nearest neighbors can be regarded as hyper-parameter which can be tuned during the learning phase of a recommendation algorithm.

A basic example similarity function which helps to figure out the similarity between user $u_a$ and $u_b$ is represented by Formula \ref{usersimilarity}. In this context, the set $F$ represents those features that have been specified, i.e., \emph{selected or deselected} by both users. 

\begin{equation}\label{usersimilarity}
    sim(u_a,u_b)=\frac{|\{f \in F: f(u_a)=f(u_b)\}|}{|\{f \in F\}|}
\end{equation}

In Table \ref{tab:usersessionsselectedfeatures}, the similarity between the current user and user $u_1$ is $1.0$, i.e., both users have completely the same preferences. In the mentioned scenario, we have to support \emph{session-based recommendation} \cite{Wang2019ASO}, since recommendations are determined by using preference information from the \emph{current user session} and the preferences of similar users. We want to emphasize that in real-world scenarios nowadays model-based machine learning approaches such as matrix factorization \cite{Koren2009} are applied for item prediction. An example of applying matrix factorization is provided in Section \ref{reconfiguration}. These approaches manage to encode complex similarity relationships into a set of \emph{hidden aspects}.

Up to now, we focused on recommending the \emph{selection or deselection of individual features}. A similar recommendation approach can also be applied to the \emph{selection of the next choice point}, i.e., which feature(s) should be presented next to the user for a selection/deselection decision. In this scenario, we recommend the next feature (the next features) a user could be interested in to specify. Note that in such scenarios information gain is often not the best criteria for attribute selection since users do not necessarily follow the criteria of information gain when selecting the next feature.

\begin{table}[ht]
\centering 
\caption{Recommendation of a possible next feature ($QA$) to be specified by the current user.}
\begin{tabular}{|c|c|c|c|c|c|c|c|c|}
\hline
  user/   &  lic  & adlic   & baslic     & AB    & stat          & QA      &bas-     &   mm-
\tabularnewline
 session   &    &    &      &     &           &       &  QA   &  QA 
\tabularnewline
\hline \hline
$u_1   $        &   1   &  3    &  2        &  4    &     5         &  6        &   8   &    7
\tabularnewline
\hline
$u_2$           &   2   &  3    &  4        &  1    &     8         &  5        &   7   &    6
\tabularnewline
\hline
$u_3$           &   1   &  2    &  3        &  5    &     8         &  4        &   7   &    6
\tabularnewline
\hline
current     &   1   &  2    &  3        &  ?    &     ?         &  ?$\rightarrow$4        &   ?   &    ?
\tabularnewline
\hline
\end{tabular}
\label{tab:userssessionfeatureordering}
\end{table}

In the example shown in Table \ref{tab:userssessionfeatureordering}, the current user has already specified his/her preferences regarding \emph{license} ($lic$), \emph{advanced license} ($adlic$), and \emph{basic license} ($baslic$). Now, we are interested in which feature the user would like to specify next. Formula \ref{userrankingsimilarity} is a variation of Formula \ref{usersimilarity} where similarity measurement focuses on the distance of feature selection orderings of two different users $u_a$ and $u_b$.

\begin{equation}\label{userrankingsimilarity}
    sim(u_a,u_b)=\frac{m(u_a,u_b) - \Sigma_{f \in F} |fr(u_a,f) - fr(u_b,f)|}{m(u_a,u_b)}
\end{equation}

 In Formula \ref{userrankingsimilarity}, $F$ denotes features that have already been specified by both users. The function $fr(u,f)$ denotes the selection position (rank) of feature $f$ in the session of user $u$.  Furthermore, the function $m$ denotes the maximum possible distance between $u_a$ and $u_b$ in terms of the order of user feature specifications. For example, $m(current,u_3)$ is $(|1-3|)+(|2-2|)+(|3-1|)$ = $2+0+2=4$ assuming the initial feature selection ordering 1:$lic$, 2:$adlic$, and 3:$baslic$. Consequently, $sim(current, u_3)$ = $\frac{4-0}{4}$ = $1.0$.
 
 After having identified the most similar user(s) of the current user, the next feature to be specified can be recommended. This is the feature with lowest ranking of the similar users that has not been specified up to now by the current user. In our example, feature $QA$ would be the recommendation candidate since the most similar user of the current user is \emph{$u_3$} and the lowest ranking of a feature not specified by the \emph{current user} is $4$ which represents $QA$ in the case of  $u_3$.
 
 \vspace{0.25cm}
 
 \textbf{Research Issues}. We want to emphasize that the mentioned recommendation approaches are not able to guarantee that the determined recommendation is \emph{consistent with the constraints in the feature model}, i.e., it could be the case that a feature setting is recommended which is \emph{inconsistent} with the underlying feature model. Basically, this means that the recommender system is unaware of the constraints defined in the feature model. A possibility to avoid such a situation is to trigger an additional consistency check before the recommendation is shown to the user. However, this requires additional computing time and could in some cases result in slow response times. An alternative is to interpret recommendations as \emph{search heuristics} (e.g., variable value orderings) and to determine recommendations (configurations) on the solver level. An initial approach to  recommendation-based search  is discussed in detail in \cite{PolatErdeniz2019}. A major challenge is to learn configurator search heuristics in such a way that a reasonable \emph{tradeoff between prediction quality and search effort} can be achieved. For related work on search optimization in feature model related reasoning we refer, for example, to Sayyad et al. \cite{Sayyad2013}.

\section{Supporting Reconfiguration}\label{reconfiguration}

With \emph{reconfiguration} \cite{Janota2014} we refer to (often interactive) scenarios where (1) a set of features has already been specified (selected or deselected) by a user but triggers an inconsistency or (2) additional features have been specified in the feature model and we would like to know ahead which user/customer is interested in extending his/her current configuration, i.e., including the new feature.

An example of the first scenario is depicted in Table \ref{tab:reconfigurationinconsistency}. The \emph{current user} has changed his/her mind and thinks about choosing a basic license ($baslic$). At the same time, it seems to be the case that the user is still interested in having an \emph{ABtesting} ($AB$) support. In this context, we have to indicate to the user which of his/her  preferences have to be adapted to restore consistency.

\begin{table}[ht]
\centering 
\caption{Log of already completed configurations (conf). The current user has specified inconsistent preferences (if $ABtesting$ is selected, a license has to be paid).}
\begin{tabular}{|c|c|c|c|c|c|c|c|c|}
\hline
  user/   &  lic  & adlic   & baslic     & AB    & stat          & QA      &bas-     &   mm-
\tabularnewline
conf   &    &    &      &     &           &       & QA    & QA  
\tabularnewline
\hline \hline
$u_1$           &   1   &  0        &  1        &  0            &     1         &  1        &   1   &    0
\tabularnewline
\hline
$u_2$           &   1   &  1        &  0        &  1            &     1         &  1        &   1   &    1
\tabularnewline
\hline
$u_3$           &   1   &  0         &  1        &  0           &     1         &  1        &   1   &    0
\tabularnewline
\hline
current     & 1     &  \bf0!     &  \bf1!     &  \bf1!            &     1         &     1     &   1   &    1
\tabularnewline
\hline
\end{tabular}
\label{tab:reconfigurationinconsistency} 
\end{table}

Technically speaking, there exist two conflicts ($CS_{1,2}$) \cite{Junker2004} between the mentioned user preferences: $CS_1: \{adlic=0,AB=1\}$ and $CS_2: \{baslic=1,AB=1\}$. This inconsistency could be resolved by pointing out to the user the option of restoring the original setting that accepts advanced licensing (alt. $1$: $adlic=1$, $baslic=0$, $AB=1$) or to accept the reduction in functionality in terms of not having available anymore  \emph{ABtesting} (alt. $2$: $adlic=0$, $baslic=1$, $AB=0$).

If more information is available about the preferences of the user, for example, we know that a user is  more interested in a simple solution ($simplicity=0.8$) compared to a full feature support ($productivity = 0.2$), we are able to rank the individual alternatives of restoring consistency. By reusing the utility scores specified in Table \ref{tab:interestdimensionsattributes}, we are able to determine a ranking for the individual change recommendations possible in the mentioned scenario (see Table \ref{tab:rankingrepairs}).

\begin{table}[ht]
\centering 
\caption{Utility evaluation of change alternatives (based on Table \ref{tab:interestdimensionsattributes}). For example, in the context of $alternative~1$ and $licensing$, $1(0.1,1)$ denotes the fact that the inclusion of licensing contributes $0.1$ to simplicity and $1.0$ to productivity.}
\begin{tabular}{|c|c|c|c|c|c|c|c|c|}
\hline
  preference    &  advanced-                  & basiclicense       & ABtesting    & utility
\tabularnewline
 change   &  license                  &        &     & 
\tabularnewline
\hline \hline
alt. 1                 &   $1(0.1,1)$     &  $0(1,0.1)$  &   $1(0.3,1)$  & 0.72
\tabularnewline
\hline
alt. 2                 &   $0(0.1,1)$       &  $1(1,0.1)$  & $0(0.3,1)$ &   0.82
\tabularnewline
\hline
\end{tabular}
\label{tab:rankingrepairs} 
\end{table}

Taking into account the preferences of our example user ($u_a$), the utility of change $alt.~1$ is $(0.1*0.8+1.0*0.2)+(0.3*0.8+1.0*0.2)=0.72$. Furthermore, the overall utility of change $alt.~2$ is $1.0*0.8+0.1*0.2=0.82$. Consequently, we could recommend change $alt.~2$ to the user which clearly focuses on the aspect of $simplicity$. Determining conflicts and corresponding change alternatives is the task of \emph{conflict detection} and corresponding \emph{diagnosis} algorithms. A discussion of these algorithms is beyond the scope of this paper. For further related details regarding conflict detection and diagnosis we refer to Junker \cite{Junker2004}, Reiter \cite{Reiter1987}, and Felfernig et al. \cite{Felfernig2012}.

Reconfiguration is not only associated with consistency management but can also be relevant when existing configurations should be extended with additional features. In Table \ref{tab:featureextension}, an additional feature \emph{share} has been added which supports the sharing of the results of a \emph{survey}. An important information in this context (e.g., in marketing scenarios) is to know which users would be interested in the new feature and should be primarily contacted. 

\begin{table}[ht]
\centering 
\caption{Session log of already completed configurations.}
\begin{tabular}{|c|c|c|c|c|c|c|c|c|c|}
\hline
  user   &  lic  & ad-   & bas-     & AB    & stat          & QA      &bas-     &   mm- & share
\tabularnewline
   &    & lic   & lic     &     &           &       &QA     &   QA & 
\tabularnewline
\hline \hline
$u_1$           &   1   &  0    &  1    &  0    &     1         &  1        &   1   &    0      & 0
\tabularnewline
\hline  
$u_2$           &   1   &  1    &  0    &  1    &     1         &  1        &   1   &    1      & 1
\tabularnewline
\hline
$u_3$           &   1   &  0    &  1    &  0    &     1         &  1        &   1   &    0      & ?
\tabularnewline
\hline
$u_4$     &   1   &  0    &  1    &  1    &     1         &  1        &   1   &    0      &     ?
\tabularnewline
\hline
\end{tabular}
\label{tab:featureextension} 
\end{table}

The \emph{relevance prediction} for a new feature for a user can be implemented with collaborative filtering (CF) \cite{Konstan1997}. A major difference compared to the previously discussed CF approaches is that the relevance of new features can be easily determined \emph{offline} which makes the task more appropriate for model-based collaborative filtering often implemented as \emph{matrix factorization} (MF) \cite{Koren2009}. MF focuses on optimizing a set of so-called hidden aspects which can then be used to predict the preferences of individual users.

Table \ref{tab:featureextension} ($T$) can be reconstructed using \emph{dimensionality reduction} which is based on the idea of learning two low-dimensional matrices ($UA$ and $AF$) that help to derive a matrix $T'$ \textasciitilde $~T$, i.e., $T'$ approximates $~T$. The advantage of this approach is generalizability and efficiency since we are able to predict individual  preferences on the basis of a simple matrix multiplication operation. Let us first construct the matrices \emph{UA} and \emph{AF} on the basis of the evaluation dimensions \emph{productivity} and \emph{simplicity} as depicted in Tables \ref{tab:matrixproductivity} and \ref{tab:matrixsimplicity}.

When applying matrix factorization to the matrices \emph{UA} and \emph{AF}, we can derive the matrix \emph{T'} which is an approximation of the original matrix \emph{T}. Note that up to now we just \emph{manually estimated} the relationship between user- and item-specific interest dimensions. The disadvantage of this approach is that estimation has to be performed manually with tedious adaptation efforts for new features and also the requirement that each user has to specify his/her preferences regarding the interest dimensions. For the new feature \emph{share} (see Table \ref{tab:matrixsimplicity}), we assume a high evaluation with regard to the dimension \emph{productivity} (additional functionality is provided) and a very high evaluation with regard to  \emph{simplicity} (the \emph{share} feature does not trigger additional complexity).

Using matrix factorization, this manual process can be substituted by machine learning that estimates the weights of individual interest dimensions to optimize the similarity between $T'$ and $T$ \cite{Koren2009}. Table \ref{tab:matrixresulting} represents the result of   a matrix multiplication \text{UA} $\bullet$ \text{AF}. In this context, \emph{share} has very high estimate ($0.94$ on a scale $0..1$) for user $u_b$. Consequently, the new feature has a high chance to be of relevance for $u_b$.

\begin{table}[ht]
\centering 
\caption{Matrix UA representing user preferences regarding interest dimensions (aspects).}
\begin{tabular}{|c|c|c|c|c|c|c|c|c|c|}
\hline
  user   &  $productivity$ & $simplicity$
\tabularnewline
\hline \hline
  $u_a$   &  0.8 & 0.2
\tabularnewline
\hline
$u_b$   &  0.2 & 0.8
\tabularnewline
\hline
\end{tabular}
\label{tab:matrixproductivity}
\end{table}

\begin{table*}[ht]
\centering 
\caption{Matrix AF representing item property (feature) relationships to interest dimensions (aspects).}
\begin{tabular}{|c|c|c|c|c|c|c|c|c|c|}
\hline
  dimension         & $adlic$   & $baslic$     & $AB$        & $stat$      &$mmQA$     &   $basQA$  & $share$
\tabularnewline
\hline \hline
$productivity$       & 1.0    &  0.1        & 1.0        & 1.0        & 1.0    & 1.0    &0.7
\tabularnewline
\hline
$simplicity$           & 0.1    &  1.0        & 0.3        & 0.5        & 0.3    & 1.0    &1.0
\tabularnewline
\hline
\end{tabular}
\label{tab:matrixsimplicity}
\end{table*}

\begin{table}[ht]
\centering 
\caption{Matrix $T'$ resulting from a matrix multiplication \text{UA} $\bullet$ \text{AF}. Feature \textsc{share} appears to be potentially relevant for $u_b$.}
\begin{tabular}{|c|c|c|c|c|c|c|c|c|c|c|}
\hline
  user         & $adlic$   & $baslic$     & $AB$        & $stat$      &$mmQA$     &   $basQA$  & $share$
\tabularnewline
\hline \hline
$u_a$       & 0.82    &  0.28        & 0.82        & 1        & 1    & 1         &0.76
\tabularnewline
\hline
$u_b$           & 0.28    &  0.82        & 0.28        & 1        & 1    & 1    &\bf0.94 
\tabularnewline
\hline
\end{tabular}
\label{tab:matrixresulting} 
\end{table}

\begin{table}[ht]
\centering 
\caption{Matrix $T'$ resulting from a matrix multiplication \text{UA} $\bullet$ \text{AF}. Feature \textsc{share} appears to be potentially relevant for $u_b$.}
\begin{tabular}{|c|c|c|c|c|c|c|c|c|c|c|}
\hline
  user         & $adlic$   & $baslic$     & $AB$        & $stat$      &$mmQA$     &   $basQA$  & $share$
\tabularnewline
\hline \hline
$u_a$       & 1    &  0        & 1        & 1        & 1    & 1         &0
\tabularnewline
\hline
$u_b$           & 0    &  1        & 0        & 1        & 1    & 1    &\bf1 
\tabularnewline
\hline
\end{tabular}
\label{tab:matrixcorrect} 
\end{table}

Such a machine learning approach substitutes manual preference specification but also has the \emph{disadvantage} that the learned dimensions (aspects) do not have a clear semantics but are just representing abstract properties that optimize the prediction quality of user interests regarding individual features. When applying matrix factorization, the interest dimensions of Tables \ref{tab:matrixproductivity} and \ref{tab:matrixsimplicity} would be substituted by two (or more) optimized abstract dimensions \cite{Koren2009}.

Many software systems are configurable in one way or another \cite{Pereira2020spaces,Pereira2020,Temple2017}. In many cases, the configuration space is huge and  mechanisms are needed that help to support tasks such as the prediction of the performance of specific (re-)configurations and the optimization of configurations. \emph{Performance prediction} can play an important role in the context of (re-)configuring packages and parameters of an operating system. In this context, it should be possible to predict system performance to avoid slow runtimes due to low-quality parametrizations.

A system should also be able to support parameter optimization, i.e., to recommend reasonable parameter settings during the ramp-up phase of a system or during reconfiguration. Such an optimization can be performed following so-called \emph{sampling, measuring, and learning} patterns \cite{Pereira2020spaces} with the overall task of identifying representative system parameter settings, measuring the impact of a specific configuration, and learn a general model to be able to classify between high- and low-quality parametrizations (configurations). 

The mentioned goals can be supported a.o. on the basis of matrix factorization \cite{Koren2009} where samples can be regarded as reference points and factorization can be applied to (1) recommend parameter settings that will result in a good system performance and (2) provide hints that some of the  parametrizations will lead to low system performance.

\vspace{0.25cm}

\textbf{Research Issues}. Similar to collaborative filtering, recommendations determined by matrix factorization cannot guarantee the feasibility of a recommendation, i.e., there can be situations where a recommendation induces an inconsistency with the constraints defined in the feature model. As mentioned, an approach to deal with such situations is to recommend solver search heuristics that are learned from existing user interaction data \cite{PolatErdeniz2019}. An important issue related to both scenarios, i.e., configuration and reconfiguration, is how to \emph{explain recommendations}. In both scenarios, explanations are shallow, i.e., only refer to the used algorithm. For example, collaborative filtering is based on explanations of the preferences of nearest neighbors. A direction for future research in this context is to combine machine learning with less-well performing knowledge-based recommendation approaches (e.g., the utility-based approach discussed in Section \ref{configurationranking}) and then generate explanations on the basis of the knowledge-based recommendation model. Such knowledge-based models allow for a more fine-grained explanation on the semantic level. An example research issue is to figure out which criteria are sufficient for the application of an inferior (in terms of prediction quality) knowledge-based approach for the generation of explanations.

\section{Supporting Modeling Processes}\label{modeling}

Finally, we take a look at the process of feature model development. In this context, recommendation approaches can be applied, for example, in the context of \emph{learning processes}. Engineers of feature models who are in charge of overtaking the development and maintenance of a new feature model, often need support in the navigation of the feature and constraint space. A basic idea to support such learning processes is to apply collaborative filtering which can help to recommend items (e.g., constraints) that could be of relevance for the knowledge engineer at a specific point of time. The recommendation approach that can be applied in this context is quite similar to the one discussed in the context of recommending features to be specified within the scope of configuration sessions. In the example depicted in Table \ref{tab:sessionspecificconstraintedits}, the current knowledge engineer ($ke$) interacting with the feature model has already visited/edited the constraints $c_1..c_3$. Collaborative recommendation can be applied to predict further relevant constraints he/she could take a look at. In our example, this would be the constraint $c_4$.

\begin{table}[ht]
\centering 
\caption{Session-specific ordering of constraint edits.}
\begin{tabular}{|c|c|c|c|c|c|c|c|c|c|}
\hline 
  session/ke   &  $c_1$& $c_2$ &  $c_3$    & $c_4$    & $c_5$    & $c_6$      &$c_7$     &   $c_8$ & $c_9$
\tabularnewline
\hline \hline
$1$           &   1   &  3    &  2        &  4    &     5         &  6        &   8   &    7 & 9
\tabularnewline
\hline
$2$           &   2   &  3    &  4        &  1    &     8         &  5        &   7   &    9 & 6
\tabularnewline
\hline
$3$           &   1   &  2    &  3        &  4    &     9         &  5        &   7   &    6 & 8
\tabularnewline
\hline
current     &   1   &  2    &  3        &  ?$\rightarrow$4    &     ?         &  ?        &   ?   &    ? & ?
\tabularnewline
\hline
\end{tabular} 
\label{tab:sessionspecificconstraintedits} 
\end{table}

Recommending the next constraint is relevant to support knowledge engineers in understanding a knowledge base. A related aspect is the \emph{grouping of constraints} in such a way that the cognitive overload of knowledge engineers can be minimized. One way to achieve this is to apply the concepts of clustering which helps to identify constraints which belong together in one way or another. An analysis of different approaches to group constraints in configuration knowledge acquisition contexts can be found, for example, in Felfernig et al. \cite{Felfernigetal2013}.

Besides supporting users in understanding a knowledge base (feature model), recommender systems can also be applied to automatically generate a knowledge base. Ulz et al. \cite{Ulz2017} introduce a \emph{human computation}\footnote{The underlying idea of \emph{human computation} is that humans take over problem solving tasks which computers are not able to solve in equal quality.} \cite{vonAhn2005} based approach where the development of recommender and configuration knowledge bases is implemented on the basis of asking users simple questions and aggregate the results in an intelligent fashion into a corresponding set of constraints. Questions refer to selection scenarios, for example, users should give feedback, which items they would like to be included in a recommendation in a specific context. The output of the approach is a set of \emph{requires} constraints that support item selection in different contexts. Bécan et al. \cite{Becan2015} and She et al. \cite{She2014} follow the similar idea of generating feature models from configuration instances, i.e., instead of asking the user questions about intended item properties, the intended properties are already represented as configurations or logical formulae.

\vspace{0.25cm}

\textbf{Research Issues}. Feature model development should be accompanied by structured testing approaches which support quality assurance of feature models, i.e., to assure that the structures and constraints defined in the feature model are consistent with the underlying domain knowledge. For example, \emph{it should not be possible} to calculate configurations that (1) include features that should not be combined with each other and (2) exclude features that are regarded as feasible in the underlying application domain. 

Quality assurance for feature models is already supported by different types of analysis operations (see, e.g., \cite{BeSeRu2010}). Open issues in this context are the \emph{recommendation of relevant test cases} useful for identifying erroneous constraints in a feature model and the recommendation of corresponding diagnoses that indicate minimal subsets of constraints which could be responsible for the faulty behavior of the feature model. Such a recommendation support would help to further improve the efficiency of feature model knowledge acquisition and maintenance processes.

\section{Conclusions and Further Research Issues}\label{conclusions}

With this paper, we provide an overview of feature modeling and configuration scenarios that can profit from the application of recommendation and related machine learning approaches. For selected scenarios, we have provided examples that help to improve the understanding of the discussed application.  Future work in this context includes a couple of empirical evaluations that will help to better estimate which recommendation approach best supports a specific scenario. There are a couple of further scenarios that could profit from the application of recommendation technologies. 

For example, the analysis of user interaction data could indicate \emph{additional relevant constraints to be included in a model}. These constraints can also be regarded as specializations of an existing knowledge base \cite{Temple2016}. This functionality could be based on association rule mining applied to the interaction data. Such an approach follows the line of research of learning whole feature models from pre-existing configurations (see, for example, Bécan et al. \cite{Becan2015}).

In knowledge acquisition scenarios, an \emph{intelligent grouping of features and constraints of a feature model} could be important to streamline maintenance processes. Supporting such a grouping can, for example, be based on clustering approaches as discussed in Felfernig et al. \cite{Felfernigetal2013}. A major issue for future work in this context is to make constraint groups flexible and adaptive to different scenarios, for example, searching for a faulty constraint or changing a specific variability property in a feature model.

A related challenge is how to recommend maintenance actions to \emph{avoid low-quality modeling} in terms of complex hierarchical structures and constraints of low understandability. Such recommendations can also be based on a more in-depth knowledge of cognitive aspects in knowledge representation and maintenance.  

Finally, an open issue in the context of applying recommendation technologies in complex item domains is how to evaluate the quality of  recommendations. Existing evaluation measures have to be adapted or extended. An  example is the measurement of \emph{precision}: in the configuration context, single features or groups of feature settings could be recommended.

\begin{acks}

The work presented in this paper has been developed within the research project \textsc{ParXCel} (\emph{Machine Learning and Parallelization for Scalable Constraint Solving}) which is funded by the Austrian Research Promotion Agency (FFG) under the project number $880657$. We want to thank the following persons for their valuable comments which helped to further improve the quality of this paper: David Benavides (University of Seville), Mayte Gómez (University of Seville), and Klaus Pilsl (Combeenation).

\end{acks}

\bibliographystyle{ACM-Reference-Format}
\bibliography{bibliography}

\end{document}